\documentclass[prl,twocolumn,showpacs,graphicx,floatfix]{revtex4}
\usepackage{epsf}

\begin{document}

\title{Universal corrections to the Fermi-liquid theory}
\author{Andrey V. Chubukov$^1$ and Dmitrii L. Maslov$^2$}
\affiliation{$^1$Department of Physics, University of Wisconsin-Madison, 1150 Univ.
Ave., Madison WI 53706-1390\\
$^2$Department of Physics, University of Florida, P. O. Box 118440, Gainesville,
FL 32611-8440}
\date{\today}
\begin{abstract}
We show that the singularities in the dynamical bosonic
response functions of a generic 2D Fermi liquid give rise to
universal, non-analytic corrections to the Fermi-liquid theory.
These corrections yield a $T^2$ term in the specific heat, $T$
terms in the effective mass and the uniform spin susceptibility
$\chi_s (Q=0,T)$, and  $|Q|$ term in $\chi_s (Q,T=0)$. The
existence of these terms has been the subject of recent
controversy, which is resolved in this paper. We present
 exact expressions
for all non-analytic terms to second order
in a generic  interaction $U(q)$ and show that only $U(0)$ and
$U(2p_F)$ matter.
\end{abstract}
 \maketitle

The universal features of a Fermi liquid and their physical
consequences continue to attract the attention of the
condensed-matter community. In a generic Fermi liquid, the
imaginary part of the retarded fermionic self-energy $\Sigma
_{R}(k,\omega )$ on the mass shell is determined solely by
fermions in a narrow $(\sim \omega $) energy range around the
Fermi surface and behaves as $\Sigma ^{\prime \prime }\propto
\omega ^{2}+(\pi T)^{2}$~\cite{agd}.
This regular behavior of the self-energy has a profound effect on
such observables as the specific heat and uniform
 spin- and charge
susceptibilities, which have the same functional dependences as
for free fermions, \emph{i.e.}, the specific heat is linear in $T$
and the susceptibilities approach finite values at $T=0$. A
regular behavior of the fermionic self-energy is also in line with
a general reasoning that turning on the interaction in $D>1$
should not affect drastically the low-energy properties of a 
 system~\cite{extra}, unless special circumstances, {\it
e.g.}, a proximity to a quantum phase transition \cite{sachdev},
interfere.

The subject of this paper is the analysis of non-analytic 
corrections to the Fermi- liquid behavior in a generic,
clean Fermi liquid. These corrections are universal in a sense
 that they are determined by fermions near the Fermi surface.  
It has been known for some time that
 corrections to the Fermi-liquid form
of $\Sigma^{\prime \prime}(\omega)$ do not form a regular,
analytic series in $\omega^{2}$ but rather scale as $\omega^{D}$
for $2 \leq D \leq 3$, with  an extra logarithm in
$D=2$~\cite{agd,baym,bloom}. [For $1<D<2$ this form persists, but it is
 not a ``correction'' anymore.] These non-analytic 
 $\omega^D$ terms (as well as non-analytic vertex corrections) 
are of
fundamental interest as they may give rise to anomalous
temperature and momentum dependences of observable quantities. A
well-known example is the $T^3\ln T$ term in the specific heat of
a 3D Fermi liquid, caused by the anomalous term
 $\Sigma^{\prime \prime}(\omega) \propto \omega^3$
  (and hence $\Sigma^\prime(\omega) \propto \omega^3 
\log \omega$)  \cite{baym}. 
A  related  example  
 is the linear-in-$T$ correction to
the conductivity  of a weakly disordered 2D system~\cite
{stern,dassarma,zna}.
Non-analytic corrections are also important for the theory of
quantum critical phenomena in itinerant ferromagnets ~\cite{qc},
as a non-analyticity of the static spin susceptibility changes
 the nature of the phase transition \cite{belitz_1}.

Belitz, Kirkpatrick and Vojta (BKV)~\cite{belitz}  and, later,
Misawa~\cite{misawa} argued that the non-analyticity in the
fermionic self-energy should gives rise to a non-analytic momentum
expansion of the particle-hole susceptibility $\chi (Q,T)$. For
non-interacting fermions, $\chi (Q,0)$ is
given by the Lindhard function and is analytic in $Q$ for small $Q$ in all $%
D $. Diagrammatically, corrections to the Lindhard function are
obtained by self-energy and vertex-corrections insertions into the
particle-hole bubble (cf. Fig.\ref {fig2}). Diagrams with
self-energy insertions are readily estimated by power counting,
and the result is that the non-analytic, $\omega ^{D}$ term in
$\Sigma $ gives rise to $\delta \chi (Q)\propto Q^{D-1}$ ,with
extra logs for $D=3$ and $D=1$. Power counting also
suggests~\cite{belitz,marenko} that the non-analyticity in $\Sigma
$ should affect the temperature dependence of the uniform
susceptibility $\chi (0,T)$ and gives rise to $\delta \chi
(0,T)\propto T^{D-1},$ with extra logarithms in $D=3$ and $D=1$.
By the same arguments, non-analyticity in $\Sigma $ should lead to
the $T^{D-1}$-dependence of the effective mass $m^* (T)$ and to
the $T^{D}$-dependence of the subleading term in the specific
heat, $\delta C(T)$,
 with extra logarithms
in 3D and 1D~\cite{doniach,bedell}.

Our motivation to pursue a further study of non-analytic
corrections to the Fermi liquid behavior is two-fold. First, we
want to verify that power counting arguments by carrying out an
explicit analytic calculations of several observable quantities:
specific heat, effective mass, and spin- and charge
susceptibilities. That power counting may be misleading is seen,
{\it e.g.}, from the example of the free-fermion susceptibility:
according to power counting, it  should also have a non-analytic
momentum dependence, whereas the exact result is analytic in $Q$
for small $Q$. The existing literature on this issue is
controversial. BKV verified their power counting arguments for the
spin susceptibility in 3D by explicitly computing $\delta \chi
_{s}(Q,T)$ to second order in the interaction. They demonstrated
that $\delta \chi_s (Q,0)\propto Q^{2}\ln |Q|$, in agreement with
power counting. At the same time,  Carneiro and Pethick
\cite{pethick} and  later
 BKV found that uniform $\chi_s (0,T)$ scales as $T^{2}$ but not as
$T^{2}\ln  T $, as predicted by power counting.  On the contrary,
Misawa did find a $T^{2}\ln T$ term in his calculation
\cite{misawa_2}. In 2D, BKV conjectured that $\chi_s (Q,0)$ scales
as $|Q|$ but no explicit calculation has not been performed.
Hirashima and Takahashi~\cite{hirashima} computed $\chi (0,T)$ in
2D numerically, but coud not draw any definite conclusions about
the $T$-dependence because of numerical difficulties.
 Chitov and Millis (CM) \cite{millis}, found analytically
 that the leading term in
$\chi_s (0,T)$ in 2D scales as $T$, in agreement with power
counting \cite{marenko}. At the same time, CM found that different
contributions to the non-analytic  $T$ term in $m^* (T)$
and  to the $T^{2}$ term in $\delta C(T)$ (both predicted
 by power counting),
cancel each other, and only analytic corrections survive.
 Meanwhile, Bedell
and Coffey~\cite{bedell} reported a $T^{2}$ term
in $\delta C(T)$. Very recently, Das Sarma et al. found a linear-in-T term
in the effective mass for the Coulomb interaction in $D=2$~\cite{dassarma_mass}.

In this paper, we present analytic results for the specific heat, effective mass,
and spin and charge susceptibilities of an interacting 2D Fermi system,
 up to second order in the short-range interaction $U(q)$. We found that
power counting arguments are generally valid, i.e., $\chi_s
(Q)\propto |Q|$, $\chi_s (T)\propto T$, $m(T)\propto T $, and
$\delta C(T)\propto T^{2}$. These results agree with Refs.
\cite{marenko,doniach,bedell}; the form of $\chi_s(T)$ agrees with
that found by CM but the forms of $m(T)$ and $\delta C(T)$ disagree
with those by CM.
Still, our prefactors for non-analytic terms differ from that
found in~\cite{bedell,millis}. We also verified that in 3D, $
\chi_s (Q,0) \propto Q^2 \ln  Q$ while  $\chi_s (0,T) \propto
T^2$, in agreement with BKV. Finally, in
 agreement with CM, we
 found no non-analytic terms in the charge susceptibility, $\chi_c$.

Another motivation for this study is to clarify the origin of the
non-analytic corrections in the Fermi-liquid theory. We found that
these corrections originate from the singularities in the
\emph{dynamic} particle-hole
response function, $\Pi (q,\Omega )$, near 
$q=0$ and $%
q=2k_{F}$, where $\Pi (q,\Omega )$ is non-analytic.
For $D=2$, near $q=0$
\begin{equation}
\Pi _{ph}^{q\approx 0}(q,\Omega _{m})=\frac{m}{2\pi }~\left( 1-\frac{|\Omega _{m}|}{%
\sqrt{(v_{F}q)^{2}+\Omega _{m}^{2}}}\right) .  \label{2.01}
\end{equation}
whereas near $q=2k_{F}$,
\begin{eqnarray}
&&\Pi _{ph}^{q \approx 2k_{F}}(q,\Omega _{m})=\frac{m}{2\pi }\times  \nonumber \\
&&\left( 1-\sqrt{\frac{\tilde{q}}{2k_F}+\left[ \left( \frac{\Omega _{m}}{%
2v_{F}k_F}\right) ^{2}+\left( \frac{\tilde{q}}{2k_F}\right) ^{2}\right]
^{1/2}}\right) ,  \label{2.02}
\end{eqnarray}
where ${\tilde{q}}\equiv q-2k_F$ and ${\tilde{q}}\ll 2k_F$.
Physically, these two singularities give rise to a zero-sound mode
and Friedel oscillations, respectively. The singularity near $q=0$
is entirely dynamic, while the one near $2k_F$ is also present in
the static limit for $q>2k_F$. We found that the singularities in
$\Pi (q, \Omega)$
 are necessary ingredients which make
 power counting arguments valid. Furthermore, we found that
 the singular pieces in the effective mass, specific heat  and $\chi
_{s}(Q,T)$ originate exclusively from the scattering amplitude
 with zero momentum transfer
\textit{and} zero total momentum: $\Gamma _{\alpha ,\beta ;\gamma ,\delta
}(k,-k;k,-k)=U(0)\delta _{\alpha \gamma }\delta _{\beta \delta
}-U(2k_F)\delta _{\alpha \delta }\delta _{\beta \gamma }$. This
implies that (i) non-analytic
terms depend only on $U(0)$ and $U(2k_F)$ but not on the interaction
averaged over the Fermi surface, and (ii)
 up to overall sign, the non-analyticities at $q \approx 0$ and
 $q \approx 2k_F$ contribute equally to
 individual diagrams for the fermionic self-energy
 and $\chi_s$, i.e.,
 singular corrections to a Fermi liquid can be  viewed {\it equivalently} as
 coming either from the singularity in the dynamical particle-hole bubble at $%
q=0 $ \textit{or} at $q=2k_{F}$. As the $q=0$ singularity is entirely
dynamical, the non-analytic corrections to a Fermi liquid are dynamical
in nature as well.

\begin{figure}[tbp]
\begin{center}
\epsfxsize=1 \columnwidth
\epsffile{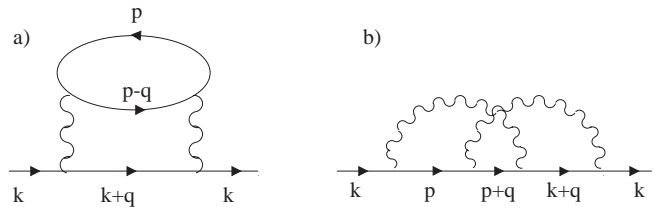}
\end{center}
\caption{Non-trivial second-order diagrams for the self-energy.}
\label{fig1}
\end{figure}

\noindent

\textit{Effective mass and specific heat.}  ~To find the effective
mass, $m^* (T)$, and the correction to the specific heat, $\delta
C(T)$, one needs to know the real part of the fermionic
self-energy, $\Sigma^\prime (k, \omega)$, on the mass shell, {\it
i.e.}, at $\epsilon_k\equiv\left(k^2-k^2_F\right)/2m=\omega$. The
two nontrivial second-order diagrams for the Fermi energy are
presented in Fig. \ref{fig1}. We evaluated $\Sigma^{\prime \prime}
(k, \omega)$ first, using the spectral representation, and then
obtained $\Sigma^\prime (\omega)$ on the mass shell via
Kramers-Kr\"{o}nig transformation. The imaginary part of the
self-energy reduces to well-known forms~\cite{bloom}
  $\Sigma^{\prime \prime }(k, \omega) \propto \omega ^{2}\ln\omega$
 and $\Sigma ^{\prime \prime }(k, \omega) \propto T^2\ln T$
for $k$ near the Fermi surface and in the limits
 of $T\to 0$ and $\omega\to 0$, respectively. We obtained $\Sigma
^{\prime \prime }(k, \omega)$ at arbitrary $\omega/T$ and
$\omega/\epsilon_k$. The full expressions are, however, rather
involved and will be presented elsewhere~\cite{ch_mas}. Using full
$\Sigma ^{\prime \prime }(k, \omega)$, we find for the real part
of the self-energy on the mass shell, neglecting a regular,
Fermi-liquid type $\omega$ term
\begin{equation}
\Sigma^{\prime }(\omega )=-\frac{m{\bar{U}}^{2}}{16\pi ^{2}v_{F}^{2}}%
\omega |\omega |g\left( \frac{\omega }{T}\right)  \label{u1.1},
\end{equation}
where ${\bar U}^2 = U^2 (0) + U^2(2p_F) - U(0) U(2p_F)$ and
\begin{equation}
g(x) = 1 + \frac{4}{x^2} \left[\frac{\pi^2}{12} + \mbox{Li}_2
\left(- e^{-|x|}\right)\right] \label{2.111}
\end{equation}
where $\mbox{Li}_2 (x)$ is a polylogarithmic function.

In the two  limits, $g(\infty) =1$ and $g(x \ll 1) \approx 4 \ln
2/x$. The first limit corresponds to $T=0$ in which case Eq.
(\ref{u1.1}) gives $\Sigma^{\prime }(\omega ) \propto \omega
|\omega|$. This non-analytic form agrees with power counting. For
small $\omega/T$, i.e., for $x \ll 1$, the $1/x$ form of $g(x)$
leads to the $\omega T$ term in  $\Sigma^{\prime }(\omega )$ for
$\omega\ll T$. This in turn implies that
 the quasiparticle mass $%
m^{\ast } (T)$ acquires a linear-in-$T$ correction
\begin{equation}
m^{\ast }(T)=m \left( 1- m^2 {\bar U}^2~ \frac{\ln  2}{8\pi^2 }~
\frac{T}{E_{F}}\right).  \label{2.11}
\end{equation}
 Using Eqs.(\ref{u1.1},\ref{2.111}), we find a correction to the specific heat
\begin{eqnarray}
\delta C(T)& =& ~\frac{2m}{T}~\left[ \int_{-\infty}^\infty d\omega
\omega \frac{\partial n}{\partial \omega }\Sigma^\prime (\omega
,\epsilon _{k}=\omega )\right] \nonumber\\
 &&= 0.174~C_{FL}~m^2 {\bar U}^2 \left( \frac{T}{E_{F}}%
\right),   \label{2.10}
\end{eqnarray}
where $C_{FL} =\pi T m/3$ is the Fermi-gas result.
 We see that a non-analyticity in
the fermionic self-energy gives rise to the $T^{2}$ term in
the specific heat. This term comes only from fermions in a near vicinity of
the Fermi surface and from this perspective is model-independent.

We also verified  that the non-analytic part of $\Sigma^{\prime}
(\omega)$
 stems exclusively from the
scattering process in which two internal momenta in the self-energy diagram are near $-%
\mathbf{k}$, whereas the third one is near $\mathbf{k}$, {\it i.e.
}, when both the total and transferred momentum are near zero.
As an independent check, we obtained
 the non-analytic part of $\Sigma^{\prime}(\omega )
$ by re-expressing the self-energy
in terms of the particle-particle
bubble--Eq. (\ref{u1.1}) then results from the (logarithmic)
singularity of the particle-particle bubble at small total momentum and
frequency. This should indeed be the case if only the scattering amplitude
 with zero total momentum is important.

\begin{figure}[tbp]
\begin{center}
\epsfxsize=0.8\columnwidth
\epsffile{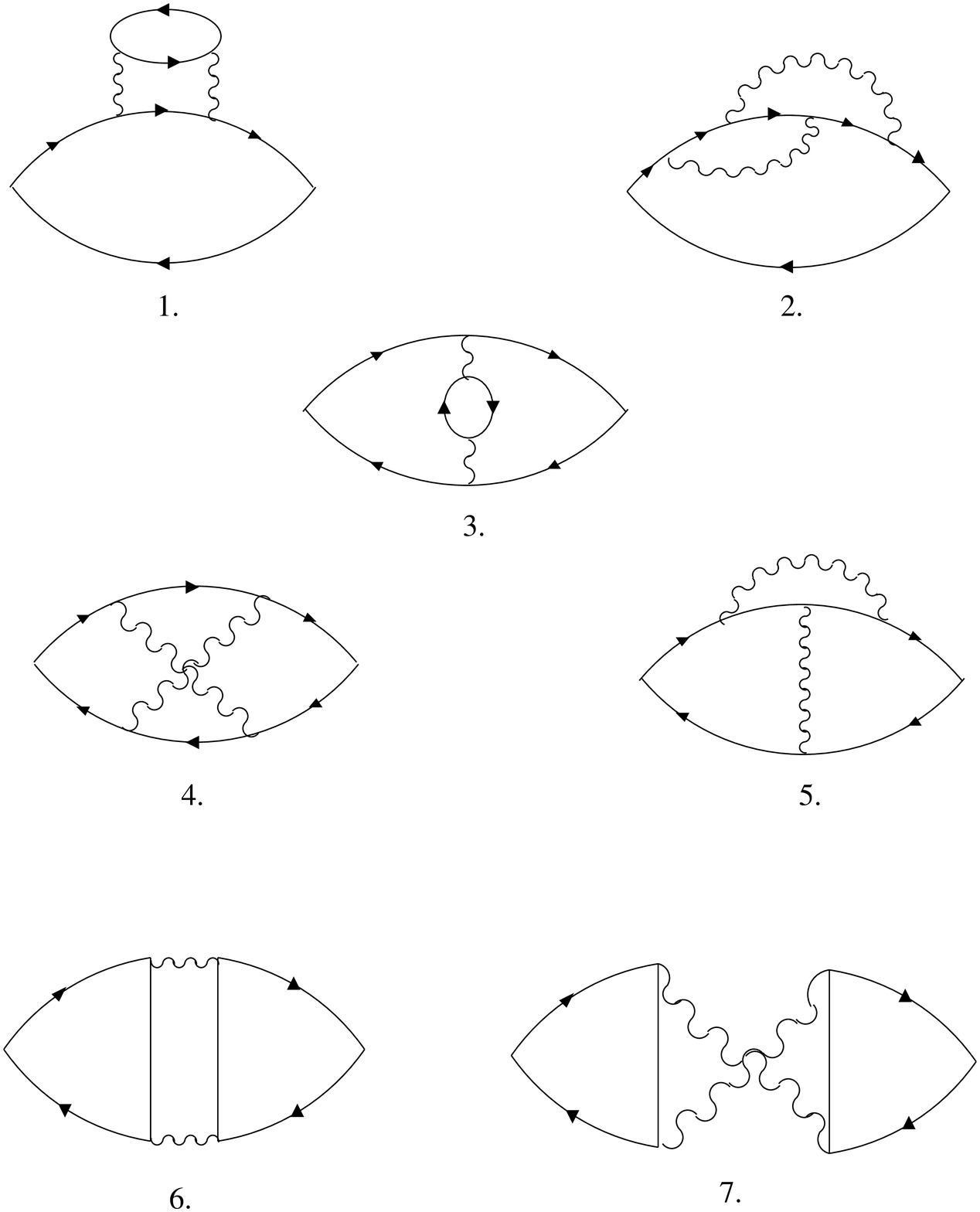}
\end{center}
\caption{Relevant second-order diagrams for spin- and charge
susceptibilities. The last two diagrams are non-zeroes only for
the charge susceptibility.} \label{fig2}
\end{figure}

\noindent \textit{Spin susceptibility.} ~ The relevant diagrams
for the spin susceptibility are presented in Fig.\ref{fig2}.
Evaluation of the diagrams is rather tedious  but straightforward.
We calculated all diagrams in two ways: i) explicitly, by
exploring the non-analyticities
 in the particle-hole bubble near $q=0$ and  $2k_F$,
 and ii) by retaining only vertices in which both total and
transferred are small. In the second approach, we expanded in
total and transferred momenta and extracted universal
contributions, which are independent of the upper cutoff for the
expansion. We obtained identical results in both methods, which
proves that only a single scattering amplitude is relevant.

The non-analytic contributions
 to the spin susceptibility from individual diagrams are as
 follows
\begin{eqnarray}
\chi _{1}(Q,T) &=&\chi _{0}K(Q,T)\left[ U^{2}(0)+U^{2}(2k_F)\right] ;
\nonumber \\
\chi _{2}(Q,T) &=&-\chi _{0}K(Q,T)U(0)U(2k_F);  \nonumber \\
\chi _{3}(Q,T) &=&\chi _{0}K(Q,T)\left[ U^{2}(2k_F)-U^{2}(0)\right] ;
\nonumber \\
\chi _{4}(Q,T) &=&\chi _{0}K(Q,T)U(0)U(2k_F); \nonumber \\
\chi _{5}&=&\chi _{6}=\chi_{7}=0,
 \label{u2}
\end{eqnarray}
where $\chi _{0}=m/\pi $, and $K(Q,0)$ and $K(0,T)$ are given by
\begin{equation}
K(Q,0)=\frac{2}{3\pi }~\left( \frac{mU}{4\pi }\right) ^{2}~\frac{|Q|}{k_F}%
;~~K(0,T)=\left( \frac{mU}{4\pi }\right) ^{2}~\frac{T}{E_{F}}.  \label{u3}
\end{equation}
Collecting all contributions, we find
\begin{equation}
\chi _{s}(Q,T)=\sum_{i=1}^{7}\chi_i(Q,T) =
 \chi _{0}~\left[1+2K(Q,T)U^{2}(2k_F)\right]. \label{u4}
\end{equation}
We see that all non-analytic contributions with $U(0)$ cancel out,
and the final result depends only on $U(2k_F)$.

We also performed a similar calculation in 3D and reproduced the
BKV's result--the analog of Eq.(\ref{u4}) but with
$K_{3D}(Q,0)=(1/18)(mk_{F}U/4 \pi ^{2})^{2}(Q/k_F)^{2}\ln  \left(
k_F/Q\right) $. In agreement with BKV, we also found that
$K(0,T)\propto T^{2}$ with no logarithmic corrections.

We now
look more deeply into how the non-analytic contributions to susceptibility
 emerge. The power-counting argument does not rely
on the singularity in the particle-hole bubble. Indeed, near
$q=0$, the singular
$\Omega _{m}/\sqrt{%
(v_{F}q)^{2}+\Omega _{m}^{2}}$  piece in $\Pi _{ph}(q,\Omega _{m})$
has the scaling dimension of
one and hence can be treated as a constant in  power counting.
 However, we found
that for each diagram, a replacement of  $\Pi
_{ph}(q,\Omega _{m})$ by a constant
does not give rise to a linear-in-$|Q|$ term in $\chi
_{s}(Q,0)$ because the prefactor
 for the $|Q|$  term
contains the integral over
$q$, in which  all poles are located in the
 same half-plane. The $q$ integral then obviously vanishes.  This vanishing could not be detected in power counting.
The substitution of the full   $\Pi _{ph}(q,\Omega _{m})$ into the
susceptibility  makes power counting arguments valid as
$\Pi _{ph}(q,\Omega _{m})$ contains a branch-cut singularity
 which   is present in both half-planes of  $q$. With this term present,
 the location of the poles in a complex $q$ plane
becomes unessential, and  the $q$ integral does not vanish.

The linear-in-$T$ dependence of $\chi (0,T)$ is also associated
with the singularity in $\Pi _{ph}(q,\Omega ),$ but the way it
emerges is
different from $T=0$ and  is similar to an anomaly.
Consider for example the $q=0$
contribution to $\chi _1(0,T)$. Leaving the integration over $q$
and summation over $\Omega _{m}$ as the last operations, we obtain
\begin{equation}
\chi _{1}^{q=0}(0,T)=-2\chi _{0}~\left( \frac{mU(0)}{4\pi }\right) ^{2}~%
\frac{T}{E_{F}}~J,  \label{c18}
\end{equation}
where
\begin{equation}
J=\sum_{m}\int dqv_{F}^{2}q~\frac{\Omega _{m}^{2}(2\Omega
_{m}^{2}-(v_{F}q)^{2})}{(\Omega _{m}^{2}+(v_{F}q)^{2})^{3}}.
 \label{c181}
\end{equation}
 Contrary to the $T=0$ case, the
 result for $J$ at finite $T$
 depends on the order of the integration and summation.
Indeed, integrating first over
$q$ in infinite limits one obtains $1/4$ for {\it all} values of $\Omega_m$,
 and a subsequent frequency summation does not
 yield a universal piece confined to low energies.
On the other hand, performing the summation over $\Omega _{m}$ first, keeping $q$
finite, and then integrating over $q$, one obtains a universal
 $-1/4$ piece in $J$ which gives rise to a linear-in-$T$ piece in $\chi_s (0,T)$. The correct way to compute $J$ is
the second one, because the $\Omega_m$-independent
 result of the integration over $q$ is inconsistent with the fact
 that the integrand in (\ref{c181})
obviously vanishes at $\Omega_m =0$ for all finite $q$. In
reality, $q$ is  always bounded from below by the inverse system
size.

We see that despite a formal analogy between the $Q$- and $T$-
dependences of $\chi_s$, the mechanism behind $\chi
_{s}(0,T)\propto T$ is very different from the one that leads to
$\chi _{s}(Q,0)\propto |Q|$ as in the latter case the order of integration is unessential.
The  factor $J$ in (\ref{c18})
can be readily found
for arbitrary $D.$ For $D\geq 2$, we have
\begin{equation}
J=-\frac{(D-2)(4-D)}{8}\left( \frac{T}{E_{F}}\right) ^{D-2}~\int_{0}^{\infty
}\frac{dzz^{D-2}}{e^{z}-1}.  \label{y3}
\end{equation}
For $D\rightarrow 2$, we reproduce $J\rightarrow
-1/4$. For
arbitrary $2< D <3$, $J \propto T^{D-2}$, i.e.,
$\chi _{s}\left( T\right) \propto T^{D-1}.$
For $D=3$, however, the integral is regular, and $J\propto T$,
i.e., $\chi _{s}(T)\propto T^{2}$ without logarithmic corrections.

\noindent
\textit{Charge susceptibility.} ~~For the charge susceptibility, we have two
additional contributuions given by diagrams 6 and 7 in Fig. 2.
We find
\begin{eqnarray}
\chi _{6}(Q,T) &=&-\chi _{0}~K(Q,T)\left[ U^{2}(0)-U^{2}(2k_F)\right] ;
\nonumber \\
\chi _{7}(Q,T) &=&-\chi _{0}~K(Q,T)\left[ U^{2}(0)+U^{2}(2k_F)\right] ,
\end{eqnarray}
where $K(Q,T)$ is given by (\ref{u3}). Combining this last result with Eq. (%
\ref{u2}), we find that all non-analytic terms from individual
diagrams cancel out, i.e., the charge susceptibility is regular,
in agreement with Ref.\cite{millis}.

To conclude, in this paper we
 demonstrated that the universal singularities
in the bosonic response functions of a Fermi liquid give rise to
universal non-analytic corrections to the Fermi-liquid forms of
the self-energy and thermodynamic variables. We obtained explicit
results in 2D for $\delta C(T)\propto T^{2}$, $\chi
_{s}(Q,T=0)\propto |Q|$, $\chi _{s}(Q=0,T)\propto T $. We
demonstrated that these non-analytic terms come from the processes
with both transferred \textit{and} total momentum close to zero.
We also demonstrated that thermal ($\propto T)$) and quantum
($\propto |Q|$) corrections to
the spin susceptibility are of different origin. This explains why in 3D, $%
\chi _{s}(Q,0)\propto Q^{2}\ln  Q$, while $\chi _{s}(0,T)\propto
T$.

We acknowledge stimulating discussions with A. Abanov, I. Aleiner,
B. Altshuler, D. Belitz, G. Chitov, A. Finkelstein, M. Mar'enko,
A. Millis, M. Norman, C. Pepin, A. Rosch, J. Rech, M. Reizer, and
Q. Si.
 The research has been supported by NSF DMR 9979749 (A. Ch.) and NSF
DMR-0077825 (DLM).

\end{document}